\documentclass[prb,twocolumn,amsmath,amssymb]{revtex4}
\usepackage[printwatermark]{xwatermark}
\usepackage{xcolor}

\usepackage{graphicx}
\usepackage{dcolumn}
\usepackage{bm}
\usepackage{epstopdf}
\def\MCO{MgCr$_2$O$_4$}

\def\E{\rm E}

\def\\beta J{\beta{\rm J}}
\def\S{\rm S}
\def\be{\begin{equation}}
\def\ee{\end{equation}}
\def\bd{\begin{displaymath}}
\def\ed{\end{displaymath}}
\def\-{\phantom{-}}
\usepackage{amsmath,amssymb}

\begin{document}

\title{Thermodynamics and spin mapping of quantum excitations in a Heisenberg spin heptamer}

\author{A. Roxburgh}
\affiliation{Department of Physics, University of North Florida, Jacksonville, FL 32224}
\author{J. T. Haraldsen}
\affiliation{Department of Physics, University of North Florida, Jacksonville, FL 32224}

\date{\today}


\begin{abstract}

In this study, we examine the thermodynamics and spin dynamics of spin-1/2 and spin-3/2 heptamers. Through an exact diagonalization of the isotropic Heisenberg Hamiltonian, we find the closed-form, analytical representations for thermodynamic properties, spin excitations, and neutron scattering structure factors. Furthermore, we investigate the  {cluster-like excitations of quantum spin heptamer} in the three-dimensional pyrochlore lattice material MgCr$_2$O$_4$. Using a spin mapping of the spin-1/2 heptamer excitations, the calculated structure factors of the spin-3/2 heptamer are be determined, which provides clarification for the spin excitations in MgCr$_2$O$_4$. Overall, this study demonstrates the ability to use the spin mapping of structure factors for small spin systems to analyze more complex structures.

\end{abstract}

\maketitle

\section{Introduction}

After over a decade of study, molecular magnets continue to intrigue researchers with new and exciting results and phenomena, which pushes towards the potential for using them in technological applications\cite{kahn:93,niel:00,godf:17,fili:18,loca:17,boga:08}. Molecular magnets are molecule-based materials that are synthesized with magnetic properties that may provide enhancements in resonant spin tunneling, quantum coherence, magnetic deflagration, and various spintronic applications\cite{bart:14,siek:17}. Typically, molecular magnets consist of clusters of magnetic spins interacting within a molecular solid that are magnetically separated from other clusters through non-magnetic ligands \cite{chie:2017,hube:17,chri:00,gatte:07,corn:06,fleu:05,nait:05,coro:04,ishi:04,boga:08,mari:18}. This has lead to the discovery of some very well investigated molecular magnetic materials, i.e. Mn$_{12}$, Ni$_{12}$, Fe$_8$, and Mn$_{84}$ to name a few. These are very large molecular magnets, which presents a challenge for an analytical analysis and modeling of the clusters.

Over the last decade, there has been an enormous theoretical thrust towards understanding the spin excitations and dynamics of clusters\cite{amed:02,efre:02,klem:02,wald:03,wald:07,ulut:03,hara:05}. While a lot of progress  {has been made} in the numerical understanding of these systems using density functional theory, $ab~initio$, and first principles studies\cite{ruiz:04,kort:03,thom:17,chie:16,rott:04,cano:07}, the ability to find and work towards straightforward closed-form solutions to experimental observables is critical to the understanding of the underlying physics in these materials.

In working towards a deeper understanding of spin clusters and magnetic systems, it has become clear that the larger spin clusters can be analyzed through the subgeometries of the cluster\cite{hara:11}. This can allow for simple, closed-form expressions for thermodynamic and magnetic properties of clusters\cite{hara:16}. Furthermore, we have shown that the magnetic ``fingerprint" of spin clusters can be determined through the inelastic neutron scattering structure, which has a functional form that is characteristic of the cluster and is independent of spin\cite{houc:15}. Therefore, understanding the nature of larger clusters lies in the analysis and characterization of the subgeometries, where total spin can be  {easily} handled.

\begin{figure}
\includegraphics[width=3.0in]{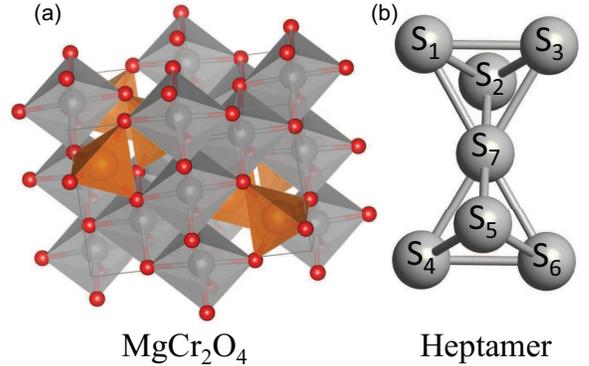}
\caption{Illustration of the \MCO~ structure (a) and the corner-sharing spin heptamer (b).}
\label{struc}
\end{figure}

While many molecular magnets are in the isolated cluster limit, some spin clusters have been observed in three-dimensional periodic lattices. One case is the discovery of a spin dimer in VODPO$_4\cdot\frac{1}{2}$D$_2$O\cite{tenn:97}. However, this has been recently observed in the pyrochlore lattice of \MCO\cite{tomi:08,tomi:13,gao:18}, where the 3D lattice forms a structure of hexagons that have corner-sharing tetrahedrons (shown in Fig. \ref{struc}(a)). This has also been observed in the spinel compound,  {ZnCr$_2$O$_4$}\cite{lee:00,lee:02,ji:09}. In the case of \MCO (shown in Fig. \ref{struc}(a)), inelastic neutron scattering studies on this compound have revealed the existence of discrete, dispersionless excitations in the energy spectra, which indicated that these are excitations of spin clusters of either six or seven spins\cite{tomi:08,tomi:13,gao:18}. 

Typically, the excitations in \MCO~ have been examined classically\cite{tomi:08,tomi:13}. However, these classical models are unable to clarify the general structure of the magnetic cluster, where they we are only able to classify them in terms of a spin hexagon and a spin heptamer (shown in Fig. \ref{struc}(b)).

In this study, we examine the spin dynamics of a quantum spin-1/2 and a spin-3/2 heptamer to help clarify the spin structure for the \MCO~ system. Using an exact diagonalization of the quantum Heisenberg Hamiltonian for a spin heptamer cluster, we show that through an analysis of the energy eigenstates, thermodynamics, and inelastic neutron scattering structure factors one can map spin excitations from a low spin cluster to higher spin cluster. Furthermore, we model the spin excitations of \MCO~ and show that the excitations observed can be described by quantum spin heptamer. This analysis was first introduced in combination with the experimental measurements and analysis presented by Gao et al.\cite{gao:18}. However, here, we show the full calculations and present the complete, closed-form representations for the energy eigenstates, thermodynamic  {properties}, and the inelastic neutron scattering structure factors.

\begin{figure}
\includegraphics[width=3.0in]{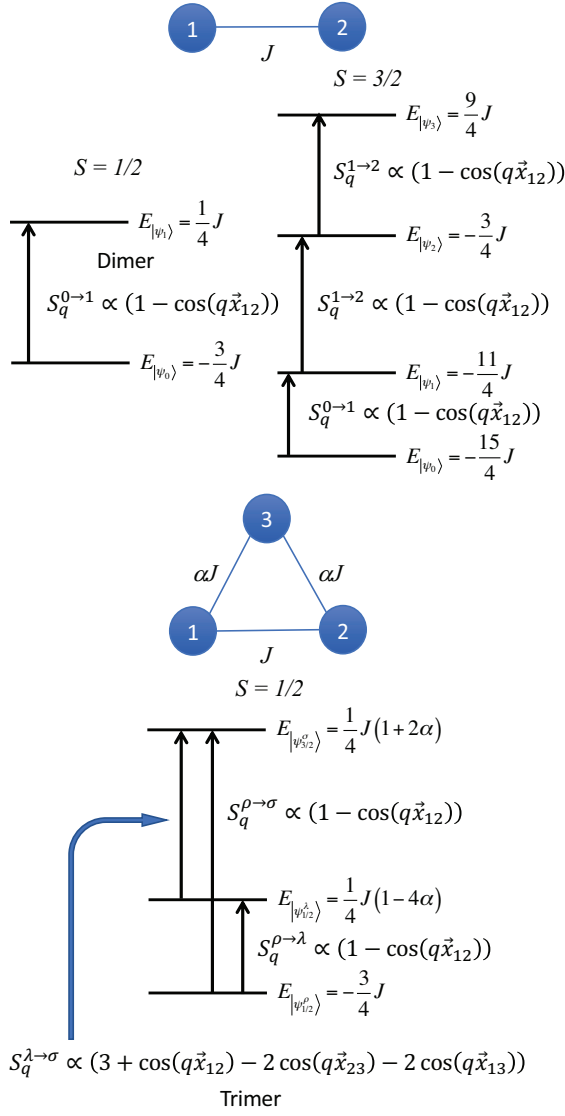}
\caption{Comparison of the spin excitations and structure factors for a $S$=1/2 dimer, trimer, and a $S$=3/2 dimer, {which shows that larger cluster excitations have characteristics of the subgeometry that is excited and that the functional form of the excitation intensity is not dependent on the spin of the system.}}
\label{DT-comp}
\end{figure}

\section{Spin Heptamer Model and Thermodynamic Properties}

The challenge in investigating large molecular magnets and spin clusters analytically is due to the size of the Hilbert space for the magnetic eignestates, where the ability to extract information from individual excitations is challenging and complicated. Therefore, to gain more information on the underlying excitations in large clusters, it is helpful to examine the trends and physics of smaller clusters to gain insight.

In recent studies, it has been shown that the underlying excitations in larger spin clusters hold on to the subgeometries of that cluster\cite{hara:05,hara:11,hara:16}. For example, in Fig. \ref{DT-comp}, we show the neutron scattering structure factor for the spin excitations for spin-1/2 dimer and a spin-1/2 isosceles trimer, and while the trimer has more excitations than the dimer, two of the excitations of the trimer consist of dimer characteristics {as shown by functional form of the embedded} dimer. Furthermore, the function form of the neutron scattering structure factor is not dependent on the total spin of the system, which has been shown in many studies\cite{houc:15}, but is illustrated in Fig. \ref{DT-comp} for the spin-3/2 dimer. These fundamental properties allow for the analytic investigation of larger spin clusters through an analysis and decomposition of the small subgeometries.

Therefore, in the case of \MCO, the system will need to be broken into smaller subgeometries in order to determine the spin excitations of the S = 3/2 heptamer. We start by examining the system using an isotropic spin-spin exchange Hamiltonian,
\be
{\cal H} = \sum_{<ij>}  {\rm J}_{ij}\;  \vec{\rm S}_{i}\cdot
\vec{\rm S}_{j},
\label{magH}
\ee
where the superexchange constants $\{ {\rm J}_{ij}\} $ are positive for antiferromagnetic interactions and negative for ferromagnetic ones, and $\vec {\rm S}_i$ is the quantum spin operator for a spin-3/2 ions at site $i$\cite{hara:05}. For a spin heptamer shown in Fig. \ref{struc}(b), the Hamiltonian can be written as 
\be 
\begin{array}{c}
\displaystyle {\cal H} = {\rm J}\, \Big\{ \Big(\vec{\rm S}_{1}\cdot\vec{\rm S}_{2} + \vec{\rm S}_{1}\cdot\vec{\rm S}_{3} + \vec{\rm S}_{2}\cdot\vec{\rm S}_{3}\Big) \\ 
\displaystyle + \Big(\vec{\rm S}_{4}\cdot\vec{\rm S}_{5} +\vec{\rm S}_{4}\cdot\vec{\rm S}_{6} + \vec{\rm S}_{5}\cdot\vec{\rm S}_{6}\Big) +  \sum_{i}^6 \vec{\rm S}_{i}\cdot \vec{\rm S}_{7} \Big\},
\end{array}
\label{heptamerH} 
\ee
where the first set of terms consists of a 1,2,3-trimer, the second set is a 4,5,6-trimer, and the final sum of terms produces a hexamer. The spin 3/2 heptamer cluster consists of 16384 total states, ($[2\cdot \frac{3}{2}+1]^7$), which makes determining every eigen-state and function analytically a challenge. While the exact diagonalization of the spin matrix can be quickly evaluated numerically, the overall functional forms for the thermodynamic properties and inelastic neutron scattering structure factors can be missed. Therefore, to evaluate the energies, we gain insight by stepping back and examining the spin-1/2 heptamer. If one considers the spin-1/2 heptamer spin states, the decomposition is given by 

\begin{figure}
\includegraphics[width=3.25in]{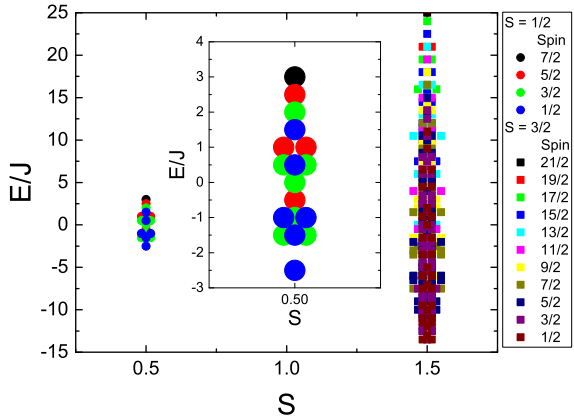}
\caption{Energy eigenstates for the spin-1/2 (inset) and spin-3/2 heptamers.}
\label{Elevels}
\end{figure}

\be
\begin{array}{ll}
{\rm Individual~ Spins} & \frac{1}{2} \otimes \frac{1}{2} \otimes \frac{1}{2} \otimes \frac{1}{2} \otimes \frac{1}{2} \otimes \frac{1}{2} \otimes \frac{1}{2} \\ \\
{\rm Trimer~ States} & \frac{1}{2} \otimes (\frac{3}{2} \oplus \frac{1}{2}^2) \otimes (\frac{3}{2} \oplus \frac{1}{2}^2) \\ \\
{\rm Hexamer~ States} & \frac{1}{2} \otimes (3 \oplus 2^{{5}} \oplus 1^{{9}} \oplus 0^{{5}}) \\ \\
{\rm Heptamer~ States} & \frac{7}{2} \oplus \frac{5}{2}^6 \oplus \frac{3}{2}^{14} \oplus \frac{1}{2}^{14}. \\
\end{array}
\ee
Here, we find the breakdown of the subgeometries of a spin hexamer produced by coupled trimer states. Here, $\otimes$ is the tensor product of the spin vector and $\oplus$ is the direct sum. The superscript indicates the multiplicity of the states. From these states, we show that when considering the magnetic degeneracy (2S + 1) the spin-1/2 heptamer consists of 128 states. When the individual spins are increased to 3/2, the number of states increases to 16384, but the spin decomposition is

\begin{widetext}
\be
\begin{array}{ll}
{\rm Individual~ Spins} & \frac{3}{2} \otimes \frac{3}{2} \otimes \frac{3}{2} \otimes \frac{3}{2} \otimes \frac{3}{2} \otimes \frac{3}{2} \otimes \frac{3}{2} \\ \\
{\rm Trimer~ States} & \frac{3}{2} \otimes (\frac{9}{2} \oplus \frac{7}{2}^2 \oplus \frac{5}{2}^3 \oplus \frac{3}{2}^4 \oplus \frac{1}{2}^2) \otimes(\frac{9}{2} \oplus \frac{7}{2}^2 \oplus \frac{5}{2}^3 \oplus \frac{3}{2}^4 \oplus \frac{1}{2}^2) \\ \\
{\rm Hexamer~ States} & \frac{3}{2} \otimes (9 \oplus 8^5 \oplus 7^{15} \oplus 6^{35} \oplus 5^{54} \oplus 4^{96} \oplus 3^{120} \oplus 2^{120} \oplus 1^{90} \oplus 0^{39}) \\ \\
{\rm Heptamer~ States} & \frac{21}{2} \oplus \frac{19}{2}^{6} \oplus \frac{17}{2}^{21} \oplus \frac{15}{2}^{56} \oplus \frac{13}{2}^{119} \oplus \frac{11}{2}^{210} \oplus \frac{9}{2}^{315} \oplus \frac{7}{2}^{400} \oplus \frac{5}{2}^{426} \oplus \frac{3}{2}^{364} \oplus \frac{1}{2}^{210}\\ \\
\end{array}
\ee
\end{widetext}
which is similar to the spin-1/2 case. Therefore, while the structure is reminiscent of two connected tetrahedrons, the energy eigenstates of this system can be determined exactly through a reduction of the full heptamer Hamiltonian using the subgeometries that are apparent in the Hamiltonian (a hexamer and two trimers basis sets). This decomposition allows us to write out the energy eigenstates for the general spin heptamer as
\be
\begin{array}{c}
E = \frac{J}{2}\Big[ S_{tot}(S_{tot}+1) - S_{hex}(S_{hex}+1) + S_{\Delta_1}(S_{\Delta_1}+1) \\ \displaystyle + S_{\Delta_2}(S_{\Delta_2}+1) - \sum_i^7 S_i(S_i+1) \Big],
\end{array}
\label{EE}
\ee
where $S_{tot}$ is the total spin state of the system, $S_{hex}$ is the hexamer spin state, $S_{\Delta_i}$ are the trimer spin states, and $S_i$ are the magnetic spins on the atoms (S = 3/2). Using the aforementioned basis sets, the eigenstates ($\big|S_{tot} S_{hex} S_{\Delta_1} S_{\Delta_2}\big>$) for the heptamer can be determined, where the eigenstates are produced due to the symmetry within the magnetic structure that produces mixing of the spin states.  {However, while the energy above is represented in hexamer and trimer components, transitions of any subgeometry (dimer, trimer, tetramer, pentamer, or hexamer) can be expected.}

\begin{figure}[b]
\includegraphics[width=3.5in]{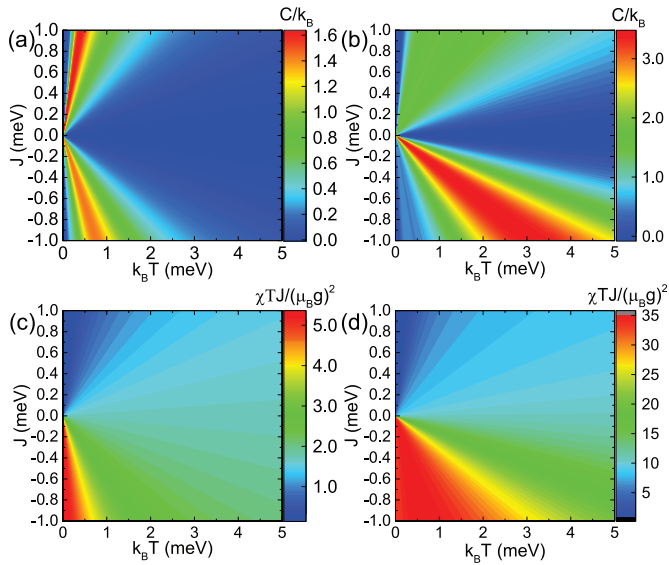}
\caption{The calculated magnetic heat capacity and magnetic susceptibility for the spin 1/2 and spin 3/2 heptamers.}
\label{CM}
\end{figure}

Figure \ref{Elevels} shows the spin-1/2 and spin-3/2 heptamer energy eigenstates, where this illustrates the dramatic increase in states with increasing spin for the heptamer structure  {and the need for spin mapping in the analytical limit.} Using these energy states, we determine the thermodynamic properties and closed-form representations for the partition function, heat capacity, and magnetic susceptibility.  {While this is a straightforward exercise, the thermodynamic properties allows for a direct check on the energy states.}

Here, we can define the partition function as
\be
 Z =  \sum_{E_i} (2S_{tot,E_i}+1)e^{-\beta E_i},
 \label{Z}
\ee
where $\beta$ = 1/$k_BT$, $E_i$ are the individual energy eigenstates,  {and $S_{tot,E_i}$ is the total spin for the eigenstate $E_i$}. From the partition function, we can examine the heat capacity by
\begin{equation}
C = k_B\beta^2\frac{d^2{\rm ln} (Z)}{d\beta^2},
\label{C}
\end{equation}
which allow us to also calculate the entropy as
\be
S = \int_0^{\infty} \frac{C}{\beta}d\beta = k_B {\rm ln}(\frac{N}{N_0}).
\ee
Here, $N$ is the total size of the Hilbert space and $N_0$ is the total number of ground states\cite{hara:05,blun:10}. Since the entropy is related to the total number of states, the integration of the heat capacity allows us to easily confirm the number of ground states and show that the energy eigenstates are correct. For the spin-1/2 heptamer, $S$ = 4$k_B$ln(2) and produces 8 total ground state levels, while the spin-3/2 heptamer produces $S$ = $k_B$ (9ln(2) - ln(3)) with 96 ground state levels, where the difference of ln(2) and ln(3) signifies a degenerate spin 1/2 and 3/2 ground states.

From the energy eigenstates, we can calculate the heat capacity as a function of temperature and $J$ (shown in Fig. \ref{CM}(a) and (b)) for both the spin-1/2 and spin-3/2 cases.  {Here, the heat capacity shows the temperature dependence of the Schottky anomaly (shift in the change of the entropy) is linear with respect to $J$ in the transition from paramagnet to antiferromagnet at low temperatures. The widening of the Schottky anomaly in the S = 3/2 heptamer is associated with increase in the number of magnetic states. While there is a change in the slope of the Schottky anomaly going from S = 1/2 to S = 3/2, the general pattern is still the same. Additionally, the change in the intensity of the heat capacity going from $J$ (antiferromagnetic) to -$J$ (ferromagnetic) is caused by change in the ground state energies, which is quite dramatic in the spin 3/2 case.}

Furthermore, the magnetic susceptibility can be determined from
\be
\chi = \frac{(g\mu_B)^2\beta}{3Z} \sum_{E_i} (2S_{tot,E_i}+1)(S_{tot,E_i}+1)S_{tot,E_i}e^{-\beta E_i},
\label{Chi}
\ee
where $g$ is the Land\'e factor and $\mu_B$ is the Bohr magneton. 

In Fig. \ref{CM}(c) and (d), we show the calculated magnetic susceptibility times temperature for the both spin cases, respectively. Most notable is the shift from antiferromagnetic low-spin ground state with a positive $J$ and the high-spin ferromagnetic with a negative interaction. {Since the systems are similar, then magnetic susceptibilities are also similar. The general broadening of the susceptibility peak in the S = 3/2 heptamer is due to the increased magnetization of the states for the ferromagnetic case.} The full closed-form, analytic representations for the spin-1/2 and spin-3/2 thermodynamics are presented in the Appendix.

\begin{figure}
\includegraphics[width=3.5in]{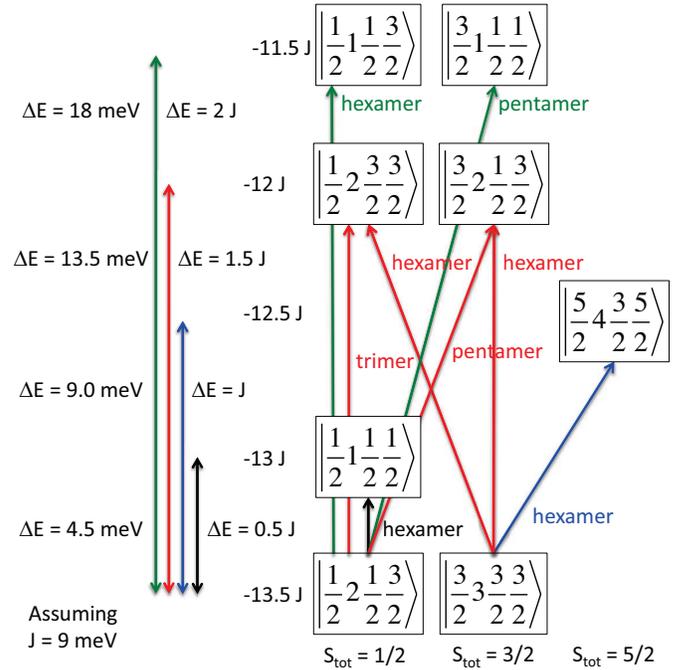}
\caption{The Heisenberg energy levels for the spin-3/2 heptamer. The states are indicated in heptamer, hexamer, and trimer basis ($\big|S_{tot} S_{hex} S_{\Delta_1} S_{\Delta_2}\big>$), where each value given the total spin for that component. On the left-hand side, the general excitation scales are given, which are in agreement with the observed spectra and estimates the superexchange energy to be about 9.0 meV. The right-hand side provides a general description of the heptamer structure considered in these calculations.}
\label{hept_energies}
\end{figure}

\begin{table*}
\caption{Structure factor Intensities for the spin heptamer}
\vskip 0.5cm
\begin{ruledtabular}
\begin{tabular}{lcc}
Excitation & Heptamer & Structure Factor \\
$\big<{\rm final}\big|{\rm initial}\big>$ & Subgeometry & Functional Form \\
\colrule
\hline
First Excitation ($\frac{1}{2}J$) &   \\
$\big<\frac{1}{2} 1 \frac{1}{2} \frac{1}{2}\big|\frac{1}{2} 2 \frac{1}{2} \frac{3}{2}\big>$ & Hexamer  & 
$3-\cos(\vec q \cdot \vec r_{23})+\cos(\vec q \cdot \vec r_{24})-\cos(\vec q \cdot \vec r_{25})-\cos(\vec q \cdot \vec r_{26})+\cos(\vec q \cdot \vec r_{27})-\cos(\vec q \cdot \vec r_{34})$\\
&& $+\cos(\vec q \cdot \vec r_{35})+\cos(\vec q \cdot \vec r_{36})-\cos(\vec q \cdot \vec r_{37})-\cos(\vec q \cdot \vec r_{45})-\cos(\vec q \cdot \vec r_{46})+\cos(\vec q \cdot \vec r_{47})$ \\ 
&& $+\cos(\vec q \cdot \vec r_{56})-\cos(\vec q \cdot \vec r_{57})-\cos(\vec q \cdot \vec r_{67})$
 \\ 
\\ \hline
Second Excitation ($J$) &   \\
$\big<\frac{5}{2} 4 \frac{3}{2} \frac{5}{2}\big|\frac{3}{2} 3 \frac{3}{2} \frac{3}{2}\big>$ & Hexamer  & 
 $6-\cos(\vec q \cdot \vec r_{23})-\cos(\vec q \cdot \vec r_{24})+\cos(\vec q \cdot \vec r_{25})+2\cos(\vec q \cdot \vec r_{26})-2\cos(\vec q \cdot \vec r_{27})+\cos(\vec q \cdot \vec r_{34})$\\
&&$ -\cos(\vec q \cdot \vec r_{35})-2\cos(\vec q \cdot \vec r_{36})+2\cos(\vec q \cdot \vec r_{37})-\cos(\vec q \cdot \vec r_{45})-2\cos(\vec q \cdot \vec r_{46})+2\cos(\vec q \cdot \vec r_{47})$\\
&& $+2\cos(\vec q \cdot \vec r_{56})-2\cos(\vec q \cdot \vec r_{57})-4\cos(\vec q \cdot \vec r_{67})$\\  
\\
\hline
Third Excitations ($\frac{3}{2}J$) &   \\
$\big<\frac{1}{2} 2 \frac{3}{2} \frac{3}{2}\big|\frac{1}{2} 2 \frac{1}{2} \frac{3}{2}\big>$ & Trimer & 
$3+\cos(\vec q \cdot \vec r_{35})-2\cos(\vec q \cdot \vec r_{37})-2\cos(\vec q \cdot \vec r_{57})$
 \\ 
 &   \\
$\big<\frac{3}{2} 2 \frac{1}{2} \frac{3}{2}\big|\frac{1}{2} 2 \frac{1}{2} \frac{3}{2}\big>$ & Pentamer\footnote{Contributes majority of intensity}  &  
 $4-2\cos(\vec q \cdot \vec r_{23})-2\cos(\vec q \cdot \vec r_{24})-2\cos(\vec q \cdot \vec r_{26})+2\cos(\vec q \cdot \vec r_{27})+\cos(\vec q \cdot \vec r_{34})+\cos(\vec q \cdot \vec r_{36})$\\
&&$ -\cos(\vec q \cdot \vec r_{37})+\cos(\vec q \cdot \vec r_{46})-\cos(\vec q \cdot \vec r_{47})-\cos(\vec q \cdot \vec r_{67})$\\
&   \\
$\big<\frac{1}{2} 2 \frac{3}{2} \frac{3}{2}\big|\frac{3}{2} 3 \frac{3}{2} \frac{3}{2}\big>$ & Hexamer$^a$  & 
$3-\cos(\vec q \cdot \vec r_{23})-\cos(\vec q \cdot \vec r_{24})+\cos(\vec q \cdot \vec r_{25})+\cos(\vec q \cdot \vec r_{26})-\cos(\vec q \cdot \vec r_{27})+\cos(\vec q \cdot \vec r_{34})$\\
&& $-\cos(\vec q \cdot \vec r_{35})-\cos(\vec q \cdot \vec r_{36})+\cos(\vec q \cdot \vec r_{37})-\cos(\vec q \cdot \vec r_{45})-\cos(\vec q \cdot \vec r_{46})+\cos(\vec q \cdot \vec r_{47})$ \\ 
&& $+\cos(\vec q \cdot \vec r_{56})-\cos(\vec q \cdot \vec r_{57})-\cos(\vec q \cdot \vec r_{67})$
\\ 
&   \\
$\big<\frac{3}{2} 2 \frac{1}{2} \frac{3}{2}\big|\frac{3}{2} 3 \frac{3}{2} \frac{3}{2}\big>$ & Hexamer  & 
$3-\cos(\vec q \cdot \vec r_{23})-\cos(\vec q \cdot \vec r_{24})-\cos(\vec q \cdot \vec r_{25})+\cos(\vec q \cdot \vec r_{26})+\cos(\vec q \cdot \vec r_{27})+\cos(\vec q \cdot \vec r_{34})$\\
&& $+\cos(\vec q \cdot \vec r_{35})-\cos(\vec q \cdot \vec r_{36})-\cos(\vec q \cdot \vec r_{37})+\cos(\vec q \cdot \vec r_{45})-\cos(\vec q \cdot \vec r_{46})-\cos(\vec q \cdot \vec r_{47})$ \\ 
&& $-\cos(\vec q \cdot \vec r_{56})-\cos(\vec q \cdot \vec r_{57})+\cos(\vec q \cdot \vec r_{67})$
\\ 
& \\
\hline
Fourth Excitations ($2J$) &   \\
$\big<\frac{1}{2} 1 \frac{1}{2} \frac{3}{2}\big|\frac{1}{2} 2 \frac{1}{2} \frac{3}{2}\big>$ & Hexamer & 
 $11+3\cos(\vec q \cdot \vec r_{23})-\cos(\vec q \cdot \vec r_{24})+\cos(\vec q \cdot \vec r_{25})-3\cos(\vec q \cdot \vec r_{26})-\cos(\vec q \cdot \vec r_{27})-3\cos(\vec q \cdot \vec r_{34})$\\
&&$ +3\cos(\vec q \cdot \vec r_{35})-9\cos(\vec q \cdot \vec r_{36})-3\cos(\vec q \cdot \vec r_{37})-\cos(\vec q \cdot \vec r_{45})+3\cos(\vec q \cdot \vec r_{46})+\cos(\vec q \cdot \vec r_{47})$\\
&& $-3\cos(\vec q \cdot \vec r_{56})-\cos(\vec q \cdot \vec r_{57})+3\cos(\vec q \cdot \vec r_{67})$\\ 
 &   \\
$\big< \frac{3}{2} 1 \frac{1}{2} \frac{1}{2}\big|\frac{1}{2} 2 \frac{1}{2} \frac{3}{2} \big>$ & Pentamer  & 
 $8+\cos(\vec q \cdot \vec r_{23})-\cos(\vec q \cdot \vec r_{24})-3\cos(\vec q \cdot \vec r_{25})+2\cos(\vec q \cdot \vec r_{27})-\cos(\vec q \cdot \vec r_{34})-3\cos(\vec q \cdot \vec r_{35})$\\
&&$ +2\cos(\vec q \cdot \vec r_{37})+3\cos(\vec q \cdot \vec r_{45})-2\cos(\vec q \cdot \vec r_{47})-6\cos(\vec q \cdot \vec r_{57})$\\
\label{SF}
\end{tabular}
\end{ruledtabular}
\end{table*}

\section{Inelastic Neutron Scattering}

Bulk probes, such heat capacity and magnetic susceptibility, are useful at determining the overall magnetic character of these clusters, but struggle to provide microscopic details of the magnetic structure. However, because of their isolation, spin clusters produce discrete dispersionless excitations that only vary in intensity when examined through neutron scattering. This allows for a ``fingerprint" of the magnetic structure through the calculations of the structure factor. 

The differential cross section for the inelastic scattering of an incident neutron from a magnetic system in an initial state $|\Psi_i\rangle$ is proportional to the neutron scattering structure factor tensor
\bd
S_{ba}(\vec q, \omega) = \hskip 3cm
\ed
\be
\int_{-\infty}^{\infty}\! \frac{dt}{2\pi} \
\sum_{\vec x{_i}, \vec x{_j}}
e^{i\vec q \cdot (\vec x_i - \vec x_j )  +i\omega t}
\langle \Psi_i |
\S_b^{\dagger}(\vec x_j, t) \S_a(\vec x_i, 0)
| \Psi_i \rangle \ .
\label{Sab_def0}
\ee
Here, the neutron has momentum transfer $\hbar\vec q$ and energy transfer $\hbar\omega$, and the site sums in Eq.(\ref{Sab_def0}) run over all magnetic ions in one unit cell, and $a,b$ are the spatial indices of the spin operators\cite{hara:05}.

For transitions between discrete energy levels, the time integral gives a trivial delta function $\delta(\E_f - \E_i - \hbar \omega)$ in the energy transfer, so it is useful to specialize to an ``exclusive structure factor" for the excitation of states within a specific magnetic multiplet (generically $|\Psi_f (\lambda_f)\rangle $) from the given initial state $|\Psi_i \rangle $,

\be
S_{ba}^{(fi)}(\vec q\, ) =
\sum_{\lambda_f}\
\langle \Psi_i |
V_b^{\dagger}
| \Psi_f (\lambda_f)\rangle \ \langle \Psi_f (\lambda_f)|
V_a
| \Psi_i \rangle  \ ,
\label{Sab_def}
\ee
where the vector $V_a(\vec q\,)$ is a sum of spin operators over all magnetic ions in a unit cell,
\be
V_a = \sum_{{\vec x}_i} {\S}_a(\vec x_i)\;
e^{i\vec q \cdot \vec x_i } \ 
\label{Va_defn}
\ee
and $a$ is the spin polarization operators $z, +, or -$\cite{squi:96}. However, in this case, since were are only interested in the functional form of the structure factor and not normalization coefficient, $V_a(\vec q\,) $ can be reduced to $V_z(\vec q\,) $ only\cite{squi:96}. Therefore, the individual structure factors for each transition can be determined.

\section{Discussion}

\subsection{Excitations in MgCr$_2$O$_4$}

The pyrochlore antiferromagnet was one of the first three dimensional models establishing that the nearest neighbor exchange does not result in magnetic ordering and that the ground state has a finite entropy\cite{chan:18}. MgCr$_{2}$O$_{4}$ is a S = 3/2 pyrochlore lattice and has been shown to display discrete magnetic clusters in a 3D periodic lattice with long-range magnetic order\cite{tomi:13}. A 3D pyrochlore lattice is generally a network of tetrahedra, whereas a 2D pyrochlore model is found by the projection of the 3D lattice on a plane. Therefore, we aim to understand the complex, high-spin magnetic interactions using spin clusters to produce discrete excitations. Previous work on this material has shown the discreteness of the magnetic excitations and have explained them using a classical spin model on hexagons and heptamer structures\cite{lee:00,lee:02,tomi:13}. However, this system has not been solved under a single consistent model on the quantum spin level.

The primary challenge in modeling the magnetic excitations is understanding the bulk unit cell in terms of magnetic clusters. One can think of the pyrochlore lattice as consisting of a magnetic structure produced through the combination of 24 heptamer structures (shown in Fig. \ref{struc}(a)). While there may be excitations of the full heptamer structure, the observed excitations from inelastic neutron scattering are likely to be excitations of the subgeometries (trimers, pentamers, hexamers, and heptamers). Therefore, the excitations of the heptamer are not restricted to that geometry, since the magnetic excitations of the larger system can take on the characteristics of the smaller subgeometries\cite{hara:11}.

Figure \ref{hept_energies} shows the lowest energy levels of the spin-3/2 heptamer, as well as their $S_{tot}$ designations and eigenstates. From this energy level diagram, it is clear that the system consists of a degenerate ground state consisting of a S = 1/2 and S = 3/2 levels. The resulting energy level diagram from the model is quite encouraging, since it reproduces the even energy spacing ($\Delta E$ = J/2) between each excitation.

\begin{figure}
\includegraphics[width=3.4in]{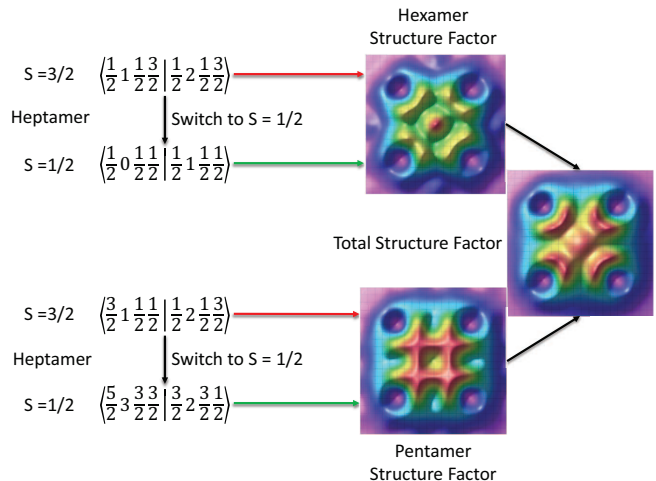}
\caption{{The general schematic for the mapping of spin excitations from the spin 1/2 heptamer to the 3/2 heptamer.}}
\label{mapping}
\end{figure}

The 4.5 meV excitation ($\Delta E$ = J/2) is a S = 1/2 to S = 1/2 transition $\big<\frac{1}{2} 1 \frac{1}{2} \frac{1}{2}\big|\frac{1}{2} 2 \frac{1}{2} \frac{3}{2}\big>$, which consists of an excitation that works through the hexamer and trimer bases. The 9.0 meV excitation ($\Delta E$ = J) is a S = 3/2 to S = 5/2 transition $\big<\frac{5}{2} 4 \frac{3}{2} \frac{5}{2}\big|\frac{3}{2} 3 \frac{3}{2} \frac{3}{2}\big>$, which also consists of an excitation through the hexamer and trimer bases. Furthermore, as shown in Fig. \ref{hept_energies}, the third and fourth excitations ($\Delta E$ = 3J/2 and $\Delta E$ = 2J, respectively) consist of a combination of multiple excitations that encompass excitations of hexamers and pentamers. It should be noted that excitations of multiple geometries are present in these transitions. This is due to the large amount of spin mixing between the heptamer, hexamer, and trimer basis sets.

 {The existence of isolated cluster-like excitation in a 3D lattice leads to further questions about the nature of interaction in materials with long-range magnetic ordering. In this case, we are not saying that a 3D pyrochlore lattice is a collection of  ``isolated” spin clusters. However, this study shows that the dispersionless excitations are similar to isolated clusters, which is in agreement with previous literature. These excitations present an exciting situation, which needs further study and comparison to different models\cite{kant:10,malk:15}.}

\subsection{Inelastic neutron scattering and the mapping of spin excitations}

\begin{figure}
\includegraphics[width=3.4in]{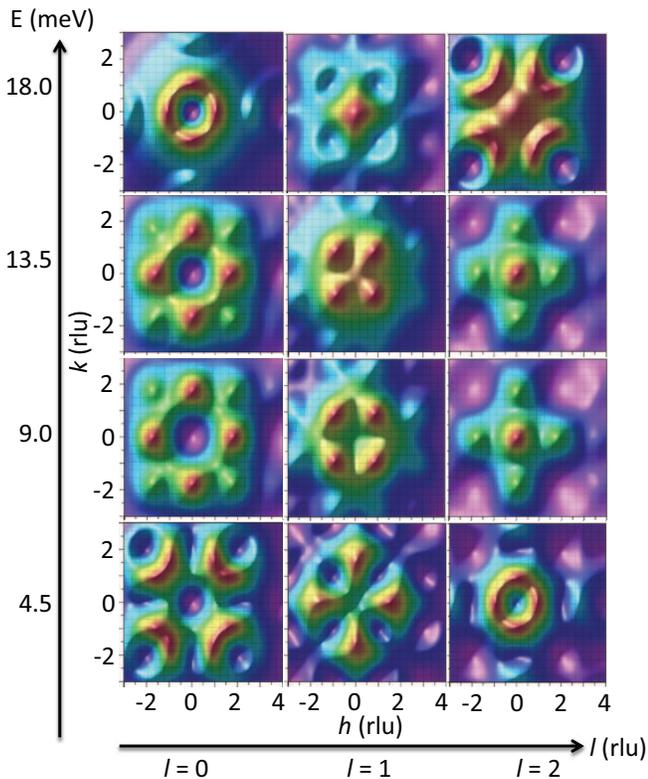}
\caption{Structure factor for the first four excitations of the spin heptamer system as a function of $h$ and $k$ for $l$ = 0, 1, and 2. These structures can be compared directly to the excitations observed in Ref. [\onlinecite{tomi:13}] and [\onlinecite{gao:18}].}
\label{SF-plot}
\end{figure}

To compare with the inelastic neutron scattering data, the single crystal structure factors have to be determined for each transitions. The spin decomposition above can help us work out the individual subgeometry basis allowing for the mapping of hexamer and trimer spin states, while the functional form of the neutron scattering structure factor can be determined by looking at individual excitations of the magnetic states. As shown in Fig. \ref{hept_energies}, the excitation from one state to another can be determined by the change in the magnetic basis. For example, the $\big<\frac{1}{2} 2 \frac{3}{2} \frac{3}{2}\big|\frac{1}{2} 2 \frac{1}{2} \frac{3}{2}\big>$ transition can be determined to be an excitation of a single trimer due to the trimer basis being the only magnetic state that changes. Therefore, that transition can be described by the standard trimer functional form. However, not all transitions are as straightforward.

Figure \ref{mapping} shows the general schematic for the mapping of spin excitations from the spin 1/2 heptamer to the 3/2 heptamer. In the case of the 18 meV excitation observed in MgCr$_2$O$_4$ and shown in Fig. \ref{hept_energies}, there are two transitions contributing to this excitation (hexamer and pentamer). Looking more closely at the $\big<\frac{1}{2} 1 \frac{1}{2} \frac{3}{2}\big|\frac{1}{2} 2 \frac{1}{2} \frac{3}{2}\big>$ transition, it is clear that characteristic transition is that of a spin hexamer. However, that structure is not known and has to be determined.

Since the sheer number of magnetic states for the S = 3/2 heptamer makes this determination very difficult to solve analytically, we have to look towards another method of identification. While the excitations can be easily determined and modeled numerically, numerical models do not provide deeper insight into the nature of interactions, whereas an analytic model can provide further information about the quantum spin states. Therefore, since the functional form of the neutron scattering structure factor is not dependent on S\cite{houc:15}, we look to reduce the spin cluster to a S = 1/2 heptamer, which consists of only 128 individual magnetic states ($(2\cdot \frac{1}{2}+1)^7$), and the analytic form of the structure factors can be determined ``easily" and compared to the data and the basis of the excitation from the S = 3/2 energy transitions.

{By looking at the similar transitions associated with the spin 1/2 heptamer, we can determine the general functional forms for transitions. Therefore, here we compared the $\big<\frac{1}{2} 0 \frac{1}{2} \frac{1}{2}\big|\frac{1}{2} 1 \frac{1}{2} \frac{1}{2}\big>$ transition of the spin 1/2 heptamer with the $\big<\frac{1}{2} 1 \frac{1}{2} \frac{3}{2}\big|\frac{1}{2} 2 \frac{1}{2} \frac{3}{2}\big>$ transition of the spin 3/2 heptamer and found the structure factors given in Table \ref{SF}. However, this was only one of the transitions available.}

{Using the same methodology, we looked into the $\big<\frac{3}{2} 1 \frac{1}{2} \frac{1}{2}\big|\frac{1}{2} 2 \frac{1}{2} \frac{3}{2}\big>$ transition. This is a challenge because the magnetic bases do not provide any direct analysis. Therefore, we compared a similar spin 1/2 heptamer transition ($\big<\frac{5}{2} 3 \frac{3}{2} \frac{3}{2}\big|\frac{3}{2} 2 \frac{3}{2} \frac{1}{2}\big>$) to the spin 3/2 transition and determined that it has characteristics of a spin pentamer. This means that only five of the seven spin sites are involved in the excitation. Once this is complete, the two structure factors are combined together to form the total structure factor, which is consistent with the data.}

Using the S = 1/2 heptamer as a basis and the methodology described above, the individual eigenfunctions for multiple transitions can be determined and compared to the transitions in the S = 3/2 case. Table \ref{SF} presents the determined functional form for the inelastic structure factor for each transition that is presented in Fig. \ref{hept_energies}. To compare to the experimental system, the individual structure factors need to be averaged over all structural configurations.

Figure \ref{SF-plot} shows the simulated inelastic neutron scattering structure factors or intensities for each transition plotted a function $h$ and $k$ for $l$ = 0, 1, 2. While the comparison is similar to that of the classical Monte Carlo simulations used in Ref. [\onlinecite{tomi:13}] and [\onlinecite{gao:18}], this model explains all transitions, including the proper energy spacing between excitations as well as the inelastic neutron scattering intensity, in a complete and self-contained model of the spin 3/2 heptamer.

As mentioned above, the individual excitation reflect the structure factors of the subgeometries of the larger magnetic structure. Since the 4.5 meV excitation is a transition between from $\big|\frac{1}{2} 2 \frac{1}{2} \frac{3}{2}\big>$ to $\big|\frac{1}{2} 1 \frac{1}{2} \frac{1}{2}\big>$ and reduces the hexamer basis from spin-2 to spin-1, the calculated structure factor is expected to be characteristic of a spin hexamer. As shown in Fig. \ref{SF-plot}, the structure factor (given in Table \ref{SF}) does reproduce the appropriate inelastic neutron scattering intensity pattern, which was originally determined to be a hexagon excitation.

Furthermore, the 9.0 meV transition is characteristic of a hexamer excitation, which is an excitation from the S = 3/2 ground state to a S = 5/2 excited state. While there is also an increase of the trimer basis, the trimer and hexamer states exhibit spin mixing due to the crossover on spin sites. The 13.5 meV and 18 meV excitations are actually a combination of multiple transitions that include mainly hexamer and pentamer excitations. While the pentamer is not a given basis, the combination of spin sites can produce excitations of five spins.

\section{Conclusion}

 {Overall, we show that the quantum spin model for clusters can be broken down to a point where the spin excitations are being produced with respect to the subgeometries of the system. Assuming a cluster can be broken into subgeometries that do not share spins, the isotropic Hamiltonian can be written analytically. This allows for the general spin energy eigenstates to be determined analytically. With the eigenstates and energies known, a mapping of excitations can be done with similar spin excitations.}

 {The subgeometries cannot share individual spins since this can lead to spin mixing that will complicate the energy eigenstates. For example, while the heptamer could be thought of as the combination of corner-sharing tetrahedrons, the sharing of only one spin will lead to mixing of the spin states. Therefore, by breaking down the heptamer into a single spin interacting with a heptamer does not mix spin states. The heptamer can then be broken into small trimer components, such that $S_{hex} = |S_{\Delta_1}-S_{\Delta_2}|$. }

 {It should be noted that the restructuring of the heptamer does not lose information about the other possible subgeometries, since the pentamers and tetramers can be formed through combinations of the other subgeometries (i.e., pentamer = trimer + dimer). This representation allows us to write down an analytical solution for the eigenstates. From the eigenstates, one can then determine the excitation ``fingerprint", which is the structure factor.}

 {Furthermore, once the energy eigenstates are determined, the thermodynamics are relatively trivial. However, they are useful for checking the ground state energies as well as illustrating the similarities and difference of the spin systems, as discussed above. Therefore, we present those solutions in full.}

In conclusion, we break down the quantum spin heptamer model and clearly show where the excitations are being produced with respect to the geometries of the system, and illustrate the ability to determine the INS excitation structure factors from the subgeometries of the spin cluster through the use of symmetry and geometry. We determine that the excitations of a small spin cluster can be mapped on to the excitations of a larger spin cluster. To do this, we examine the thermodynamics and inelastic neutron scattering structure factors for a spin-1/2 and-3/2 heptamer and provide analytical solutions for all observables. The modeled spin excitations were further compared to the inelastic neutron scattering excitations observed in MgCr$_2$O$_4$\cite{lee:00,lee:02,tomi:13,gao:18}, where the model excitations are in good agreement with the observed measurements. Therefore, these calculations help clarify the spin excitations and solidify the magnetic structure for these excitations. We believe this will help in the understanding of the larger single molecular magnets by providing a methodology for a simple and easy way to identify quantum spin excitations in magnetic clusters.

\section*{Acknowledgements}

A.R. and J.T.H acknowledge support from the Institute for Materials Science at Los Alamos National Laboratory, which provided undergraduate research support. Furthermore, we thank O. Zaharko for useful discussions on the \MCO~ pyrochlore excitations.

\section{Appendix A}

\subsection{Thermodynamics for the spin-$\frac{1}{2}$ heptamer}

While the spin=1/2 has 128 individual magnetic states, many of them are degenerate in the heptamer configuration we are considering. From Eq. \ref{EE} and Fig. \ref{Elevels}, the energy eigenstates can be determined exactly. Therefore, the partition function for the spin-1/2 heptamer is determined in a simple closed-form representation from Eq. \ref{Z} and given as
\be
\begin{array}{ll}
Z_{1/2} = &8\,{{\rm e}^{-3\,\beta J}}+6\,{{\rm e}^{\frac{1}{2}\,\beta J}}+6\,{{\rm e}^{-\frac{5}{2}\,\beta J}}+
24\,{{\rm e}^{-\beta J}}+4\\
& +24\,{{\rm e}^{\frac{3}{2}\,\beta J}}+4\,{{\rm e}^{-2\,\beta J}}+18
\,{{\rm e}^{-\frac{1}{2}\,\beta J}}+24\,{{\rm e}^{\beta J}}\\
&+8\,{{\rm e}^{\frac{5}{2}\,\beta J}}+2\,{
{\rm e}^{-\frac{3}{2}\,\beta J}}
\end{array}
\ee
From Eqs. \ref{C} and \ref{Chi}, we can determine the heat capacity and magnetic susceptibility for the spin-1/2 heptamer to be
\be
\begin{array}{ll}
\frac{C_{1/2}}{k_B} = & \frac{\beta^{2}{J}^{2}{\rm e}^{\frac{1}{2}\,\beta J}}{2}\, \Big( 96\,{\rm e}^{19
/2\,\beta J}+216\,{{\rm e}^{9\,\beta J}}+96\,{{\rm e}^{\frac{17}{2}\,\beta J}}\\
&+ 172\,{{\rm e}^{8\,\beta J}}+720\,{{\rm e}^{\frac{15}{2}\,\beta J}}+1302\,{{\rm e}^{7\,\beta J}}+1040\,{
{\rm e}^{\frac{13}{2}\,\beta J}}\\
&+2613\,{{\rm e}^{6\,\beta J}}+2022\,{{\rm e}^{\frac{11}{2}\,\beta J}}
+1877\,{{\rm e}^{5\,\beta J}}+1656\,{{\rm e}^{\frac{9}{2}\,\beta J}}\\
&+2964\,{{\rm e}^{4\,
\beta J}}+1748\,{{\rm e}^{\frac{7}{2}\,\beta J}}+456\,{{\rm e}^{3\,\beta J}}+408\,{{\rm e}^{5
/2\,\beta J}}\\
&+ 613\,{{\rm e}^{2\,\beta J}}+390\,{{\rm e}^{\frac{3}{2}\,\beta J}}+21\,{{\rm e}^
{\beta J}}+16\,{{\rm e}^{\frac{1}{2}\,\beta J}}+6 \Big)  \\
&\Big/ \Big( 12\,{{\rm e}^{4\,\beta J}
}+3\,{{\rm e}^{\frac{7}{2}\,\beta J}}+3\,{{\rm e}^{\frac{1}{2}\,\beta J}}+12\,{{\rm e}^{\frac{9}{2}\,\beta J}
}+9\,{{\rm e}^{\frac{5}{2}\,\beta J}}\\
&+4\,{\rm e}^{\frac{11}{2}\,\beta J}+{\rm e}^{\frac{3}{2}\,\beta J}+2\,{\rm e}^{3\,\beta J}+12\,{\rm e}^{2\,\beta J}+2\,{\rm e}^{\beta J}+4\Big)^{2}
\end{array}
\ee
and
\be
\begin{array}{ll}
\chi_{1/2} = & \frac{\beta(g\mu_B)^2}{2Z}\, \Big( 84\,{{\rm e}^{-3\,\beta J}}+35\,{{\rm e}^{\frac{1}{2}\,\beta J}}+
35\,{{\rm e}^{-\frac{5}{2}\,\beta J}}+140\,{{\rm e}^{-\beta J}}\\
&+10+44\,{{\rm e}^{\frac{3}{2}\,\beta J
}}+10\,{{\rm e}^{-2\,\beta J}}+41\,{{\rm e}^{-\frac{1}{2}\,\beta J}}+44\,{{\rm e}^{\beta J}}\\
&+4\,{{\rm e}^{\frac{5}{2}\,\beta J}}+{{\rm e}^{-\frac{3}{2}\,\beta J}} \Big)
\end{array}
\ee
respectively. This allows us to present the heat capacity and the magnetic susceptibility as a function of $J$ and $T$ in Fig. \ref{CM}.

\subsection{Thermodynamics for the spin-$\frac{3}{2}$ heptamer}

Similar to the spin-1/2 heptamer, the spin-3/2 heptamer has many degenerate states. Therefore, even though it has 16384 individual magnetic states, the thermodynamics can be determined in closed-form. Here, the partition function is given by
\begin{widetext}
\be
\begin{array}{ll}
Z_{3/2} = &\Big( 20+18\,{{\rm e}^{\frac{3}{2}\,\beta J}}+94\,{{\rm e}^{\frac{9}{2}\,\beta J}}+16\,{
{\rm e}^{3\,\beta J}}+74\,{{\rm e}^{\frac{15}{2}\,\beta J}}+22\,{{\rm e}^{\frac{1}{2}\,\beta J}}+84\,
{{\rm e}^{6\,\beta J}}+126\,{{\rm e}^{\frac{19}{2}\,\beta J}}+140\,{{\rm e}^{\frac{21}{2}\,\beta J}}+
10\,{{\rm e}^{\frac{23}{2}\,\beta J}}\\&
+84\,{{\rm e}^{9\,\beta J}}+16\,{{\rm e}^{10\,\beta J}}+
90\,{{\rm e}^{{\frac {25}{2}}\,\beta J}}+108\,{{\rm e}^{11\,\beta J}}+216\,{
{\rm e}^{{\frac {27\,\beta J}{2}}}}+116\,{{\rm e}^{12\,\beta J}}+4\,{{\rm e}^{13
\,\beta J}}+144\,{{\rm e}^{14\,\beta J}}+66\,{{\rm e}^{{\frac {29\,\beta J}{2}}}}\\&
+244
\,{{\rm e}^{15\,\beta J}}+306\,{{\rm e}^{{\frac {31\,\beta J}{2}}}}+48\,{{\rm e}
^{16\,\beta J}}+236\,{{\rm e}^{{\frac {33\,\beta J}{2}}}}+252\,{{\rm e}^{17\,\beta J}
}+30\,{{\rm e}^{{\frac {35\,\beta J}{2}}}}+348\,{{\rm e}^{18\,\beta J}}+246\,{
{\rm e}^{{\frac {37\,\beta J}{2}}}}+144\,{{\rm e}^{19\,\beta J}}\\&
+386\,{{\rm e}^{
{\frac {39\,\beta J}{2}}}}+240\,{{\rm e}^{20\,\beta J}}+234\,{{\rm e}^{{\frac {
41\,\beta J}{2}}}}+392\,{{\rm e}^{21\,\beta J}}+264\,{{\rm e}^{{\frac {43\,\beta J}{2
}}}}+276\,{{\rm e}^{22\,\beta J}}+410\,{{\rm e}^{{\frac {45\,\beta J}{2}}}}+312
\,{{\rm e}^{23\,\beta J}}+318\,{{\rm e}^{{\frac {47\,\beta J}{2}}}}\\&
+536\,{
{\rm e}^{24\,\beta J}}+480\,{{\rm e}^{{\frac {49\,\beta J}{2}}}}+176\,{{\rm e}^{
25\,\beta J}}+226\,{{\rm e}^{{\frac {51\,\beta J}{2}}}}+552\,{{\rm e}^{26\,\beta J}}+
334\,{{\rm e}^{{\frac {53\,\beta J}{2}}}}+304\,{{\rm e}^{27\,\beta J}}+678\,{
{\rm e}^{{\frac {55\,\beta J}{2}}}}+292\,{{\rm e}^{28\,\beta J}}\\& +518\,{{\rm e}^{
{\frac {57\,\beta J}{2}}}}+468\,{{\rm e}^{29\,\beta J}}+144\,{{\rm e}^{{\frac {
59\,\beta J}{2}}}}+568\,{{\rm e}^{30\,\beta J}}+414\,{{\rm e}^{{\frac {61\,\beta J}{2
}}}}+308\,{{\rm e}^{31\,\beta J}}+694\,{{\rm e}^{{\frac {63\,\beta J}{2}}}}+360
\,{{\rm e}^{32\,\beta J}}+244\,{{\rm e}^{{\frac {65\,\beta J}{2}}}}\\&
+560\,{
{\rm e}^{33\,\beta J}}+288\,{{\rm e}^{{\frac {67\,\beta J}{2}}}}+244\,{{\rm e}^{
34\,\beta J}}+344\,{{\rm e}^{{\frac {69\,\beta J}{2}}}}+312\,{{\rm e}^{35\,\beta J}}+
222\,{{\rm e}^{{\frac {71\,\beta J}{2}}}}+256\,{{\rm e}^{36\,\beta J}}+216\,{
{\rm e}^{{\frac {73\,\beta J}{2}}}}+120\,{{\rm e}^{37\,\beta J}}\\&
+96\,{{\rm e}^{{
\frac {75\,\beta J}{2}}}}+192\,{{\rm e}^{38\,\beta J}}+8\,{{\rm e}^{{\frac {77\,
\beta J}{2}}}}+96\,{{\rm e}^{39\,\beta J}} \Big) {{\rm e}^{-{\frac {51\,\beta J}{2}
}}}.
\end{array}
\ee
\end{widetext}

From the partition, we can then calculate the heat capacity from Eq. \ref{C}, which will be

\begin{widetext}
\be
\begin{array}{ll}
\frac{C_{3/2}}{k_B} = &\frac{\beta^{2}{J}^{2}{\rm e}^{\frac{1}{2}\,\beta J}}{2}\, \Big( 55+127514303\,{{\rm e}^
{53\,\beta J}}+133332070\,{{\rm e}^{{\frac {103\,\beta J}{2}}}}+129946362\,{
{\rm e}^{54\,\beta J}}+198\,{{\rm e}^{\frac{3}{2}\,\beta J}}+16544\,{{\rm e}^{\frac{9}{2}\,\beta J}}+
405\,{\rm e}^{\beta J}\\
&+1100\,{\rm e}^{3\,\beta J}+39886\,{\rm e}^{\frac{15}{2}\,\beta J}+19359\,{\rm e}^{4\,\beta J}+27951\,{\rm e}^{6\,\beta J}+128266\,{\rm e}^{\frac{19}{2}\,\beta J}+58626\,{\rm e}^{7\,\beta J}+1440\,{\rm e}^{\frac{5}{2}\,\beta J}\\
&+357256\,{{\rm e}^{\frac{21}{2}\,\beta J}}
+337904\,{{\rm e}^{\frac{23}{2}\,\beta J}}+37854\,{{\rm e}^{11/
2\,\beta J}}+180474\,{{\rm e}^{9\,\beta J}}+233630\,{{\rm e}^{10\,\beta J}}+164592\,{
{\rm e}^{{\frac {25}{2}}\,\beta J}}\\&
+154606\,{{\rm e}^{11\,\beta J}}+98064\,{
{\rm e}^{\frac{17}{2}\,\beta J}}+985140\,{{\rm e}^{{\frac {27\,\beta J}{2}}}}+439687\,{
{\rm e}^{12\,\beta J}}+665576\,{{\rm e}^{13\,\beta J}}+111879160\,{{\rm e}^{{
\frac {109\,\beta J}{2}}}}\\&
+47345748\,{{\rm e}^{{\frac {123\,\beta J}{2}}}}+
48228378\,{{\rm e}^{61\,\beta J}}+124424586\,{{\rm e}^{44\,\beta J}}+153718608\,
{{\rm e}^{{\frac {101\,\beta J}{2}}}}+177621180\,{{\rm e}^{48\,\beta J}}\\&
+
115286082\,{{\rm e}^{{\frac {111\,\beta J}{2}}}}+24047532\,{{\rm e}^{64\,\beta J
}}+24159384\,{{\rm e}^{{\frac {129\,\beta J}{2}}}}+21500973\,{{\rm e}^{65\,
\beta J}}+15625842\,{{\rm e}^{{\frac {131\,\beta J}{2}}}}\\&
+33179499\,{{\rm e}^{63
\,\beta J}}+29666310\,{{\rm e}^{{\frac {127\,\beta J}{2}}}}+126743778\,{{\rm e}^
{{\frac {107\,\beta J}{2}}}}+63882174\,{{\rm e}^{{\frac {119\,\beta J}{2}}}}+
17717163\,{{\rm e}^{66\,\beta J}}\\&
+12224000\,{{\rm e}^{{\frac {133\,\beta J}{2}}}
}+150237240\,{{\rm e}^{{\frac {89\,\beta J}{2}}}}+66911956\,{{\rm e}^{59\,b
J}}+80153376\,{{\rm e}^{{\frac {117\,\beta J}{2}}}}+138793727\,{{\rm e}^{47
\,\beta J}}\\&
+80621274\,{{\rm e}^{58\,\beta J}}+40967568\,{{\rm e}^{62\,\beta J}}+
158872215\,{{\rm e}^{49\,\beta J}}+133180682\,{{\rm e}^{{\frac {91\,\beta J}{2}}
}}+80210430\,{{\rm e}^{{\frac {115\,\beta J}{2}}}}\\&
+159454635\,{{\rm e}^{46
\,\beta J}}+52198392\,{{\rm e}^{{\frac {121\,\beta J}{2}}}}+97932114\,{{\rm e}^{
57\,\beta J}}+97586160\,{{\rm e}^{{\frac {113\,\beta J}{2}}}}+10250076\,{{\rm e}
^{67\,\beta J}}\\&
+142482504\,{{\rm e}^{{\frac {105\,\beta J}{2}}}}+10675668\,{
{\rm e}^{{\frac {135\,\beta J}{2}}}}+460738\,{{\rm e}^{14\,\beta J}}+1315032\,{
{\rm e}^{{\frac {29\,\beta J}{2}}}}+1846244\,{{\rm e}^{15\,\beta J}}+1017598\,{
{\rm e}^{{\frac {31\,\beta J}{2}}}}\\&
+1792833\,{{\rm e}^{16\,\beta J}}+2474800\,{
{\rm e}^{{\frac {33\,\beta J}{2}}}}+1037196\,{{\rm e}^{17\,\beta J}}+3026202\,{
{\rm e}^{{\frac {35\,\beta J}{2}}}}+3762438\,{{\rm e}^{18\,\beta J}}+1855624\,{
{\rm e}^{{\frac {37\,\beta J}{2}}}}\\&
+5081706\,{{\rm e}^{19\,\beta J}}+5765296\,{
{\rm e}^{{\frac {39\,\beta J}{2}}}}+2955929\,{{\rm e}^{20\,\beta J}}+7272576\,{
{\rm e}^{{\frac {41\,\beta J}{2}}}}+7574206\,{{\rm e}^{21\,\beta J}}+4289360\,{
{\rm e}^{{\frac {43\,\beta J}{2}}}}\\&
+10452450\,{{\rm e}^{22\,\beta J}}+10010704\,
{{\rm e}^{{\frac {45\,\beta J}{2}}}}+7130288\,{{\rm e}^{23\,\beta J}}+15103806\,
{{\rm e}^{{\frac {47\,\beta J}{2}}}}+14540404\,{{\rm e}^{24\,\beta J}}+167558226
\,{{\rm e}^{{\frac {99\,\beta J}{2}}}}\\&
+135444944\,{{\rm e}^{{\frac {97\,\beta J
}{2}}}}+11262224\,{{\rm e}^{{\frac {49\,\beta J}{2}}}}+16892797\,{{\rm e}^{
25\,\beta J}}+17444030\,{{\rm e}^{{\frac {51\,\beta J}{2}}}}+16534990\,{{\rm e}^
{26\,\beta J}}+21994040\,{{\rm e}^{{\frac {53\,\beta J}{2}}}}\\&
+25258778\,{{\rm e}
^{27\,\beta J}}+22950800\,{{\rm e}^{{\frac {55\,\beta J}{2}}}}+30834737\,{
{\rm e}^{28\,\beta J}}+36048736\,{{\rm e}^{{\frac {57\,\beta J}{2}}}}+23635886\,
{{\rm e}^{29\,\beta J}}+34749918\,{{\rm e}^{{\frac {59\,\beta J}{2}}}}\\&
+47193317
\,{{\rm e}^{30\,\beta J}}+31435088\,{{\rm e}^{{\frac {61\,\beta J}{2}}}}+
45967663\,{{\rm e}^{31\,\beta J}}+62635262\,{{\rm e}^{{\frac {63\,\beta J}{2}}}}
+39104651\,{{\rm e}^{32\,\beta J}}+60461360\,{{\rm e}^{{\frac {65\,\beta J}{2}}}
}\\&
+72937255\,{{\rm e}^{33\,\beta J}}+46361532\,{{\rm e}^{{\frac {67\,\beta J}{2}}
}}+73517486\,{{\rm e}^{34\,\beta J}}+78659680\,{{\rm e}^{{\frac {69\,\beta J}{2}
}}}+60606664\,{{\rm e}^{35\,\beta J}}+91611918\,{{\rm e}^{{\frac {71\,\beta J}{2
}}}}\\&
+88557950\,{{\rm e}^{36\,\beta J}}+75631944\,{{\rm e}^{{\frac {73\,\beta J}{
2}}}}+105700731\,{{\rm e}^{37\,\beta J}}+104147808\,{{\rm e}^{{\frac {75\,b
J}{2}}}}+86113090\,{{\rm e}^{38\,\beta J}}+110028928\,{{\rm e}^{{\frac {77
\,\beta J}{2}}}}\\&
+120588501\,{{\rm e}^{39\,\beta J}}+144069768\,{{\rm e}^{52\,\beta J}
}+94823872\,{{\rm e}^{56\,\beta J}}+170737899\,{{\rm e}^{45\,\beta J}}+111072168
\,{{\rm e}^{55\,\beta J}}\\&
+35053272\,{{\rm e}^{{\frac {125\,\beta J}{2}}}}+
62471592\,{{\rm e}^{60\,\beta J}}+175504272\,{{\rm e}^{{\frac {93\,\beta J}{2}}}
}+157451766\,{{\rm e}^{{\frac {95\,\beta J}{2}}}}+136513754\,{{\rm e}^{50\,
\beta J}}\\&
+158477088\,{{\rm e}^{51\,\beta J}}+94655378\,{{\rm e}^{{\frac {79\,\beta J
}{2}}}}+6735120\,{{\rm e}^{{\frac {137\,\beta J}{2}}}}+120513408\,{{\rm e}^
{40\,\beta J}}+138875400\,{{\rm e}^{{\frac {81\,\beta J}{2}}}}\\&
+108028132\,{
{\rm e}^{41\,\beta J}}+132388290\,{{\rm e}^{{\frac {83\,\beta J}{2}}}}+156038637
\,{{\rm e}^{42\,\beta J}}+114686736\,{{\rm e}^{{\frac {85\,\beta J}{2}}}}+
3671646\,{{\rm e}^{70\,\beta J}}\\&
+142546647\,{{\rm e}^{43\,\beta J}}+6184731\,{
{\rm e}^{69\,\beta J}}+6577786\,{{\rm e}^{68\,\beta J}}+160085898\,{{\rm e}^{{
\frac {87\,\beta J}{2}}}}+3599808\,{{\rm e}^{{\frac {139\,\beta J}{2}}}}+2767344
\,{{\rm e}^{{\frac {141\,\beta J}{2}}}}\\&
+1477856\,{{\rm e}^{71\,\beta J}}+1618728
\,{{\rm e}^{{\frac {143\,\beta J}{2}}}}+972966\,{{\rm e}^{72\,\beta J}}+642368\,
{{\rm e}^{{\frac {145\,\beta J}{2}}}}+513744\,{{\rm e}^{73\,\beta J}}+356280\,{
{\rm e}^{{\frac {147\,\beta J}{2}}}}\\&
+185032\,{{\rm e}^{74\,\beta J}}+125568\,{
{\rm e}^{{\frac {149\,\beta J}{2}}}}+68184\,{{\rm e}^{75\,\beta J}}+23424\,{
{\rm e}^{{\frac {151\,\beta J}{2}}}}+10560\,{{\rm e}^{76\,\beta J}}+9216\,{
{\rm e}^{{\frac {153\,\beta J}{2}}}}+96\,{{\rm e}^{77\,\beta J}} \Big)\\&
\Big/ \Big( 10+9\,{{\rm e}^{\frac{3}{2}\,\beta J}}+47\,{{\rm e}^{\frac{9}{2}\,\beta J}}+8\,{{\rm e}^
{3\,\beta J}}+37\,{{\rm e}^{\frac{15}{2}\,\beta J}}+11\,{{\rm e}^{\frac{1}{2}\,\beta J}}+42\,{{\rm e}
^{6\,\beta J}}+63\,{{\rm e}^{\frac{19}{2}\,\beta J}}+70\,{{\rm e}^{\frac{21}{2}\,\beta J}}+5\,{
{\rm e}^{\frac{23}{2}\,\beta J}}\\&
+42\,{{\rm e}^{9\,\beta J}}+8\,{{\rm e}^{10\,\beta J}}+45\,{
{\rm e}^{{\frac {25}{2}}\,\beta J}}+54\,{{\rm e}^{11\,\beta J}}+108\,{{\rm e}^{{
\frac {27\,\beta J}{2}}}}+58\,{{\rm e}^{12\,\beta J}}+2\,{{\rm e}^{13\,\beta J}}+72\,
{{\rm e}^{14\,\beta J}}+33\,{{\rm e}^{{\frac {29\,\beta J}{2}}}}+122\,{{\rm e}^{
15\,\beta J}}\\&
+153\,{{\rm e}^{{\frac {31\,\beta J}{2}}}}+24\,{{\rm e}^{16\,\beta J}}+
118\,{{\rm e}^{{\frac {33\,\beta J}{2}}}}+126\,{{\rm e}^{17\,\beta J}}+15\,{
{\rm e}^{{\frac {35\,\beta J}{2}}}}+174\,{{\rm e}^{18\,\beta J}}+123\,{{\rm e}^{
{\frac {37\,\beta J}{2}}}}+72\,{{\rm e}^{19\,\beta J}}+193\,{{\rm e}^{{\frac {39
\,\beta J}{2}}}}\\&
+120\,{{\rm e}^{20\,\beta J}}+117\,{{\rm e}^{{\frac {41\,\beta J}{2}}
}}+196\,{{\rm e}^{21\,\beta J}}+132\,{{\rm e}^{{\frac {43\,\beta J}{2}}}}+138\,{
{\rm e}^{22\,\beta J}}+205\,{{\rm e}^{{\frac {45\,\beta J}{2}}}}+156\,{{\rm e}^{
23\,\beta J}}+159\,{{\rm e}^{{\frac {47\,\beta J}{2}}}}+268\,{{\rm e}^{24\,\beta J}}\\&
+
240\,{{\rm e}^{{\frac {49\,\beta J}{2}}}}+88\,{{\rm e}^{25\,\beta J}}+113\,{
{\rm e}^{{\frac {51\,\beta J}{2}}}}+276\,{{\rm e}^{26\,\beta J}}+167\,{{\rm e}^{
{\frac {53\,\beta J}{2}}}}+152\,{{\rm e}^{27\,\beta J}}+339\,{{\rm e}^{{\frac {
55\,\beta J}{2}}}}+146\,{{\rm e}^{28\,\beta J}}+259\,{{\rm e}^{{\frac {57\,\beta J}{2
}}}}\\&
+234\,{{\rm e}^{29\,\beta J}}+72\,{{\rm e}^{{\frac {59\,\beta J}{2}}}}+284\,
{{\rm e}^{30\,\beta J}}+207\,{{\rm e}^{{\frac {61\,\beta J}{2}}}}+154\,{{\rm e}^
{31\,\beta J}}+347\,{{\rm e}^{{\frac {63\,\beta J}{2}}}}+180\,{{\rm e}^{32\,\beta J}}
+122\,{{\rm e}^{{\frac {65\,\beta J}{2}}}}+280\,{{\rm e}^{33\,\beta J}}\\&
+144\,{
{\rm e}^{{\frac {67\,\beta J}{2}}}}+122\,{{\rm e}^{34\,\beta J}}+172\,{{\rm e}^{
{\frac {69\,\beta J}{2}}}}+156\,{{\rm e}^{35\,\beta J}}+111\,{{\rm e}^{{\frac {
71\,\beta J}{2}}}}+128\,{{\rm e}^{36\,\beta J}}+108\,{{\rm e}^{{\frac {73\,\beta J}{2
}}}}+60\,{{\rm e}^{37\,\beta J}}+48\,{{\rm e}^{{\frac {75\,\beta J}{2}}}}\\&
+96\,{
{\rm e}^{38\,\beta J}}+4\,{{\rm e}^{{\frac {77\,\beta J}{2}}}}+48\,{{\rm e}^{39
\,\beta J}} \Big) ^{2}.
\end{array}
\ee
\end{widetext}
Lastly, using Eq. \ref{Chi}, the magnetic susceptibility can be written as
\begin{widetext}
\be
\begin{array}{ll}
\chi_{3/2} = &\,\frac {\beta}{2{\it Z}} {{\rm e}^{-{\frac {51\,\beta J}{2}}}} \Big( 1330+969\,{{\rm e}^{\frac{3\beta J}{2}\,}}+5775
\,{{\rm e}^{\frac{9\beta J}{2}\,}}+680\,{{\rm e}^{3\,\beta J}}+2885\,{{\rm e}^{\frac{15\beta J}{2}\,}
}+1771\,{{\rm e}^{\frac{\beta J}{2}\,}}+4162\,{{\rm e}^{6\,\beta J}}\\
&+6783\,{\rm e}^{19/2\,\beta J}+5685\,{\rm e}^{\frac{21\beta J}{2}\,}+165\,{\rm e}^{\frac{23\beta J}{2}\,}+3234\,
{\rm e}^{9\,\beta J}+680\,{{\rm e}^{10\,\beta J}}+2940\,{{\rm e}^{{\frac {25}{2
}}\,\beta J}}+4366\,{{\rm e}^{11\,\beta J}}\\
&+7596\,{{\rm e}^{{\frac {27\,\beta J}{2}}}
}+3474\,{{\rm e}^{12\,\beta J}}+10\,{{\rm e}^{13\,\beta J}}+176\,{{\rm e}^{{
\frac {75\,\beta J}{2}}}}+5160\,{{\rm e}^{38\,\beta J}}+4\,{{\rm e}^{{\frac {77
\,\beta J}{2}}}}+176\,{{\rm e}^{39\,\beta J}}+5592\,{{\rm e}^{14\,\beta J}}\\
&+2721\,{
{\rm e}^{{\frac {29\,\beta J}{2}}}}+7454\,{{\rm e}^{15\,\beta J}}+10504\,{
{\rm e}^{{\frac {31\,\beta J}{2}}}}+744\,{{\rm e}^{16\,\beta J}}+5270\,{{\rm e}^
{{\frac {33\,\beta J}{2}}}}+7090\,{{\rm e}^{17\,\beta J}}+495\,{{\rm e}^{{\frac 
{35\,\beta J}{2}}}}+8326\,{{\rm e}^{18\,\beta J}}\\
&+7131\,{{\rm e}^{{\frac {37\,\beta J
}{2}}}}+4440\,{{\rm e}^{19\,\beta J}}+11180\,{{\rm e}^{{\frac {39\,\beta J}{2}}}
}+4468\,{{\rm e}^{20\,\beta J}}+6351\,{{\rm e}^{{\frac {41\,\beta J}{2}}}}+9300
\,{{\rm e}^{21\,\beta J}}+5280\,{{\rm e}^{{\frac {43\,\beta J}{2}}}}+5794\,{
{\rm e}^{22\,\beta J}}\\
&+8525\,{{\rm e}^{{\frac {45\,\beta J}{2}}}}+9184\,{{\rm e}
^{23\,\beta J}}+6841\,{{\rm e}^{{\frac {47\,\beta J}{2}}}}+8888\,{{\rm e}^{24\,b
J}}+9708\,{{\rm e}^{{\frac {49\,\beta J}{2}}}}+4572\,{{\rm e}^{25\,\beta J}}+
4305\,{{\rm e}^{{\frac {51\,\beta J}{2}}}}+11420\,{{\rm e}^{26\,\beta J}}\\
&+7022\,
{{\rm e}^{{\frac {53\,\beta J}{2}}}}+4056\,{{\rm e}^{27\,\beta J}}+12509\,{
{\rm e}^{{\frac {55\,\beta J}{2}}}}+5022\,{{\rm e}^{28\,\beta J}}+8934\,{{\rm e}
^{{\frac {57\,\beta J}{2}}}}+8634\,{{\rm e}^{29\,\beta J}}+3432\,{{\rm e}^{{
\frac {59\,\beta J}{2}}}}+9300\,{{\rm e}^{30\,\beta J}}\\
&+6495\,{{\rm e}^{{\frac {
61\,\beta J}{2}}}}+6474\,{{\rm e}^{31\,\beta J}}+11662\,{{\rm e}^{{\frac {63\,\beta J
}{2}}}}+7680\,{{\rm e}^{32\,\beta J}}+4686\,{{\rm e}^{{\frac {65\,\beta J}{2}}}}
+7848\,{{\rm e}^{33\,\beta J}}+4608\,{{\rm e}^{{\frac {67\,\beta J}{2}}}}+5290\,
{{\rm e}^{34\,\beta J}}\\
&+6296\,{{\rm e}^{{\frac {69\,\beta J}{2}}}}+5484\,{
{\rm e}^{35\,\beta J}}+4586\,{{\rm e}^{{\frac {71\,\beta J}{2}}}}+3224\,{{\rm e}
^{36\,\beta J}}+3612\,{{\rm e}^{{\frac {73\,\beta J}{2}}}}+812\,{{\rm e}^{37\,\beta J
}} \Big) .
\end{array}
\ee
\end{widetext}
The advantage of writing out these formulas is to illustrate the increased complexity in going from a spin 1/2 to a spin 3/2 heptamer.

\end{document}